\documentclass[aps,prd,superscriptaddress,onecolumn,showpacs,10pt]{revtex4-1}

\usepackage{graphicx,color}
\usepackage{bm}
\usepackage{amssymb,amsmath,amsfonts, mathrsfs}
\usepackage{natbib}
\usepackage {indentfirst}
\usepackage{float}

\usepackage{ulem} 

\newcommand{\be}{\begin{equation}}
\newcommand{\ee}{\end{equation}}
\newcommand{\ba}{\begin{array}}
\newcommand{\ea}{\end{array}}
\newcommand{\bqa}{\begin{eqnarray}}
\newcommand{\eqa}{\end{eqnarray}}

\newcommand{\ket}{\rangle}
\newcommand{\bra}{\langle}

\begin{document}


\title{Scrutinizing $\pi\pi$ scattering in light of recent lattice phase shifts }


\author{Xiu-Li Gao}
\affiliation{School of Physics, Southeast University, Nanjing 211189,
P.~R.~China}
\author{Zhi-Hui Guo}
\email[]{zhguo@seu.edu.cn}
\affiliation{School of Physics, Southeast University, Nanjing 211189,
P.~R.~China}
\author{Zhiguang Xiao}
\email[]{xiaozg@scu.edu.cn}
\affiliation{School of Physics, Si Chuan University, Chengdu~610065,
P.~R.~China}
\author{Zhi-Yong Zhou}
\email[]{zhouzhy@seu.edu.cn}
\affiliation{School of Physics, Southeast University, Nanjing 211189,
P.~R.~China}


\date{\today}

\begin{abstract}
In this paper, the $IJ=00, 11, 20$ partial wave $\pi\pi$ scattering
phase shifts determined by the lattice QCD approach are analyzed by
using a  novel dispersive solution of the S-matrix, i.e. the PKU representation, in which the unitarity and
analyticity of scattering amplitudes are automatically satisfied and the phase shifts are conveniently decomposed into the contributions of the cuts and various poles, including bound states, virtual states and resonances.
The contribution of the left-hand cut is estimated by the $SU(2)$
chiral perturbation theory to $\mathcal{O}(p^4)$. The
Balanchandran-Nuyts-Roskies relations are considered as constraints to
meet the requirements of the crossing symmetry. It is found that the
$IJ=00$ $\pi\pi$ scattering phase shifts obtained at $m_\pi=391$~MeV
by Hadron Spectrum Collaboration (HSC) reveal the presence of both a
bound state pole and a virtual state pole below the $\pi\pi$ threshold
rather than only one bound state pole for the $\sigma$. To reproduce
the lattice phase shifts at $m_\pi=391$~MeV, a virtual-state pole in
the $IJ=20$ channel is found to be necessary  in order to balance the
left-hand cut effects from the chiral amplitudes. Similar discussions
are also carried out for the lattice results with $m_\pi=236$~MeV from
HSC. The observed behaviors of the pole positions with respect to the
variation of the pion masses can provide deep insights into our
understanding of the dynamical origin of $\sigma$ resonance.
\end{abstract}


\maketitle
\section{Introduction}
The $\pi\pi$ scattering is one of the most fundamental and important
process in hadron physics, and it has been investigated for long to
understand the non-perturbative aspects of quantum
chromodynamics~(QCD). The near threshold behaviors of $\pi\pi$ scatterings
have been precisely described by the chiral perturbation
theory~($\chi$PT)~\cite{Gasser:1983yg}. Moreover, the resonances,
appearing as the intermediate states of $\pi\pi$ scatterings,
have attracted
much interest from experimental, phenomenological and lattice QCD communities. The
$\pi\pi$ phase shift in the $IJ=00$ channel rises smoothly from the
elastic threshold and reaches $\pi/2$ at about 1.0 GeV, which does not
exhibit a typical resonance
lineshape~\cite{Grayer:1974cr,Rosselet:1976pu,Pislak:2003sv}. Such an
observation raises a long time debate about whether the $\sigma$
resonance exists or not until the $\sigma$ pole is simultaneously determined by several model-independent approaches such as Roy equation or those dispersive methods respecting the unitarity, analyticity, and the crossing symmetry of the scattering
amplitude~\cite{Caprini:2005zr,Zhou:2004ms,Garcia-Martin:2011nna,Yao:2020bxx,Pelaez:2015qba}. For further details, see the mini review of scalar resonances provided by PDG~\cite{ParticleDataGroup:2020ssz}.

The  $\sigma$ state has been studied for more than six decades because
it plays an important role in the nucleon nucleon interaction or in
the dynamics of spontaneous breaking of chiral symmetry. Many
different models are proposed to understand its nature but the only
consensus of different groups might be that the $\sigma$ is not a
conventional $q\bar q$ states. In the last seventies, it was proposed
that the light scalar meson states might be described by a tetraquark
model which shows an inverted mass relation compared with the
conventional quark model~\cite{Jaffe:1976ig,Maiani:2004uc}. A unitarized meson model was developed in Refs.~\cite{vanBeveren:1982qb,vanBeveren:1986ea}, and it is concluded that the $\sigma$ and other light scalar states could be dynamically generated in that model. Alternatively, it was suggested in Ref.~\cite{Tornqvist:1995kr} that a bare scalar $q\bar q$ seed when coupled to the two-meson channels with the same quantum numbers, could generate an additional set of dynamical states, corresponding to the light scalar nonet below 1~GeV, apart from the seed states, corresponding to the ones above 1~GeV~\cite{Boglione:1997aw,Zhou:2010ra}. Such a mechanism
could be better understood in a relativistic Friedrichs-Lee model since there are light $u$ and $d$ quarks involved, and the scheme could be also extended to study the states with heavy $c$ or $b$
quarks~\cite{zhou:2020,Zhou:2020moj}.  Another method to study the
$\sigma$ resonance is to unitarize the $\chi$PT amplitudes with some
unitarization scheme and extract the poles of the scattering
amplitude~\cite{ollerPhysRevLett.80.3452,Guo:2011pa}. The $N_C$ trajectories of resonance poles are also helpful to
discriminate the nature of the resonances~\cite{Pelaez:2006nj,Guo:2021blc}.

Recently, the lattice QCD groups have developed new techniques so
that all the required quark propagation diagrams, including the
quark-antiquark annihilation ones, could be evaluated with good
statistical precision. Such unquenched lattice simulations mean
that the scalar isoscalar states, which has the same quantum numbers as the vacuum, are able to be studied from the first principle
method~\cite{Briceno:2016mjc,Dudek:2012gj,Wilson:2015dqa}. Due to the
limitation of computing resources, the pion mass employed in the current lattice QCD calculation is still larger than its physical value. For example, several
calculations of the $\pi\pi$ scatterings with $IJ=00, 11, 20$ have
been performed in
Refs.~\cite{Briceno:2016mjc,Dudek:2012gj,Wilson:2015dqa} with
$m_\pi=236$ and 391 MeV. Those results with unphysically larger pion
masses, which cannot be produced in the experiments,  provide a different useful insight into the internal features of the hadron resonances.

In this paper, we extract the information of various poles from the
phase shift data of $IJ=00, 11, 20$ $\pi\pi$ scattering provided by  lattice QCD. We make use of the novel dispersive solution of the S-matrix, namely the PKU representation~\cite{zheng:2003rw}, to form
the partial wave scattering amplitude, in which the left-hand cut is
calculated from the $\chi$PT up to $O(p^4)$. To restore the crossing symmetry of the $\pi\pi$ scattering amplitudes, the Balanchandran-Nuyts-Roskies~(BNR)
relations~\cite{Balachandran:1968zza,Roskies:1969pe} are used as penalty functions to constrain the partial wave
amplitudes of different channels. We find that the
lattice data of the $IJ=00$ $\pi\pi$ scattering phase shifts with $m_\pi=391$~MeV prefer a senario with both a bound-state pole and a nearby virtual-state one rather than the case with only one bound state in a na\"ive fit.
Such a difference is helpful in
distinguishing the proposed models and shedding more light on the
nature of $\sigma$ state. Furthermore, a virtual
state pole in $IJ=20$ $\pi\pi$ scattering channel is found to be needed in order to properly reproduce the lattice phase shifts.

The paper is organized as follows. In Section \ref{theory}, the
formalism adopted in the analysis is introduced. The PKU
parametrization form is introduced in section~\ref{PKUform}. The
$SU(2)$ $\chi$PT $\pi\pi$ amplitude to $O(p^4)$ is briefly summarized
in section~\ref{chPTsect}. The BNR relations of partial $s$ and $p$
wave amplitudes are presented in section~\ref{bnr}.
Section~\ref{numerical} is devoted to the numerical analyses and
discussions. The last section is left for the summary and conclusions.

\section{Theoretical background}\label{theory}

\subsection{PKU representation}\label{PKUform}
There are two types of singularities of the scattering amplitudes: poles and cuts. By
factorizing the contributions of a bound state pole, a virtual state
pole, or a pair of conjugated resonance poles, the simplest scattering
$S$ matrix could be solved from the generalized unitarity relation of
partial wave amplitudes~\cite{zheng:2003rw}. Then, the partial-wave two-body scattering $S$-matrix
can be decomposed as multiplicative forms of the contributions from poles and the cuts~\cite{zheng:2003rw}
\begin{align*}\label{eq.smat}
S(s)=S_{cut}(s)\,\prod_v S_v(s)\prod_b S_b(s)\prod_r S_r(s)\,,
\end{align*}
where the subscripts $cut$, $v$, $b$ and $r$ stand for the effects of the two-body continuum cuts, virtual states, bound states and resonances, respectively. The explicit forms of the $S$-matrix solutions $S_{i=v,b,r}$ have been worked out in ref.~\cite{zheng:2003rw}. For the sake of completeness, we simply give their expressions in this work. The virtual state, with its pole at $s_v$ below the threshold on the real axis of second Riemann sheet of complex $s$-plane, gives
\begin{align}
S_{v}(s)=\frac{1+i\rho(s)\frac{s}{s-s_L}\sqrt{\frac{s_v-s_L}{s_R-s_v}}}{1-i\rho(s)\frac{s}{s-s_L}\sqrt{\frac{s_v-s_L}{s_R-s_v}}}\,.
\end{align}
For a bound state, with its pole at $s_b$ below the threshold on the real axis of the first Riemann sheet, its expression reads
\begin{align}
S_{b}(s)=\frac{1-i\rho(s)\frac{s}{s-s_L}\sqrt{\frac{s_b-s_L}{s_R-s_b}}}{1+i\rho(s)\frac{s}{s-s_L}\sqrt{\frac{s_b-s_L}{s_R-s_b}}}\,.
\end{align}
For a resonance, with a pair of conjugated poles $z_r$ and $z_r^*$ on the second Riemann sheet, its contribution to the $S$-matrix is given by
\begin{align}
S_r(s)=\frac{M^2[z_r]-s+i\rho(s)sG_r}{M^2[z_r]-s-i\rho(s)sG_r}\,,
\end{align}
where
\begin{align}
M^2_r&=\mathrm{Re}[z_r]+\mathrm{Im}[z_r]\frac{\mathrm{Im}[\sqrt{(z_r-s_R)(z_r-s_L)}]}{\mathrm{Re}[\sqrt{(z_r-s_R)(z_r-s_L)}]},\nonumber\\
G_r&=\frac{\mathrm{Im}[z_r]}{\mathrm{Re[\sqrt{(z_r-s_R)(z_r-s_L)}]}},
\end{align}
and the kinematic factor is expressed as
\bqa
\rho(s)=\sqrt{\frac{(s-s_L)(s-s_R)}{s^2}},
\eqa
with $s_L=(m_1-m_2)^2$ and $s_R=(m_1+m_2)^2$. Here the masses of the two scattering particles are $m_1$ and $m_2$. For the equal-mass case studied in the $\pi\pi$ scattering, the formula could be simplified, due to the neat relations $s_L=0$ and $s_R=4m_\pi^2$.

Comparing with various pole effects, $S_{cut}$, contributed by the left-hand cuts~($l.h.c.$) and the inelastic right-hand cuts~($r.h.c.$) starting from the first inelastic threshold, can be regarded as the background term in Eq.~\eqref{eq.smat}. Its explicit form can be parameterized as~\cite{zheng:2003rw}
\begin{align}\label{scut}
S_{cut}=e^{2i\rho(s)f(s)}\,,
\end{align}
where $f(s)$ is given by a once-subtracted dispersion relation
\begin{align}
f(s)=f(s_0)+\frac{s-s_0}{\pi}\int_{L}\frac{\text{Im}_L f(s')}{(s'-s_0)(s'-s)}ds'+\frac{s-s_0}{\pi}\int_{R}\frac{\text{Im}_R f(s')}{(s'-s_0)(s'-s)}ds'\,,
\end{align}
with $L$ the $l.h.c.$, $R$ the $r.h.c$ from the first inelastic threshold to $\infty$, and $s_0$ the subtraction point. It is
usually chosen at $s_0=0$ so that the asymptotic behaviour leads to a
vanishing subtraction constant~\cite{Zhou:2004ms}. For $\pi\pi$
scattering, the first inelastic threshold $4\pi$ is weakly coupled, and
the $K\bar K$ threshold is far away from the $\pi\pi$ threshold, so
the $r.h.c.$ plays a minor role in determining the $\sigma$ resonance
and it is omitted in the numerical analysis. Thus, the main
contribution to $f(s)$ comes from the left-hand cut
\begin{align}\label{eq.lhcf}
f(s)=\frac{s}{\pi}\int_{-\infty}^{0}\frac{\text{Im}_L ~f(s')}{s'(s'-s)}ds',
\end{align}
in which the imaginary part of $f(s)$ can be estimated by the $\chi$PT
amplitude as illustrated in section~\ref{chPTsect}.

By construction, the PKU representation automatically satisfies the partial wave unitarity
and the analyticity of the scattering amplitudes required by the micro-causality condition. Since it is solved from the generalized unitarity
relation of $S$-matrix with minimal assumptions, the representation is
rigorous for two body scatterings. Furthermore, the factorization of the cuts and various poles
leads to a novel feature of the phase shifts from the cuts
and poles, which makes it much convenient and powerful when analyzing the
phase-shift data without being troubled by the contribution of possible
``spurious" poles generated by some unitarization approaches~\cite{Qin:2002hk}. The PKU representation has been successfully applied to extracting the pole positions of $\sigma$ and
$\kappa$~\cite{zheng:2003rw,Zhou:2004ms}, and later on it is also extended to the $\pi N$ scattering and predicts the novel $N^*(890)$ resonance in $S_{11}$ channel~\cite{Wang:2017agd,Yao:2020bxx}.

\subsection{$\pi\pi$ scattering amplitudes of $\chi$PT and estimation
of the left-hand cut integral}\label{chPTsect}

It is rather challenging to reliably calculate the left-hand
integral in eq.~\eqref{eq.lhcf}, since it lies in the unphysical region. In
this work, we use the chiral effective field theory to estimate the
integral along the left-hand cut. Since we focus on the elastic $\pi\pi$ scattering below the $K\bar{K}$
threshold in this work, it is enough to use the $SU(2)$ chiral
Lagrangians in our study.  We stick to the conventional $SU(2)$ $\chi$PT up to
one-loop level, for which the amplitudes include the $O(p^2)$
tree-level terms, the terms from the $O(p^4)$ local operators and the ones from the one-loop diagrams generated by the $O(p^2)$ interactions.
The relevant $O(p^2)$ and $O(p^4)$ $\chi$PT
Lagrangians for the $\pi\pi$ scattering are respectively given by
\begin{eqnarray}
\mathcal{L}_{2}= \frac{F^2}{4} \bra u^\mu u_\mu + \chi_+ \ket\,,
\end{eqnarray}
and
\begin{eqnarray} \mathcal{L}_{4}=  \frac{l_1}{4} \bra u_\mu u^\mu \ket \bra u_\nu u^\nu \ket + \frac{l_2}{4}\bra u^\mu u^\nu \ket \bra  u_\mu u_\nu \ket
+ \frac{l_3+l_4}{16} \bra \chi_+ \ket \bra \chi_+ \ket
+ \frac{l_4}{8} \bra u_\mu u^\mu \ket \bra \chi_+ \ket \,.
\end{eqnarray}

We use the following convention to perform the partial-wave projection
of the $\pi\pi$ scattering amplitude
\begin{eqnarray}
 T_{IJ}= \frac{1}{2} \frac{1}{s-4m_\pi^2} \int_{4m_\pi^2-s}^{0} T_{I}(s,t) P_J(1+\frac{2t}{s-4m_\pi^2}) \,,
\end{eqnarray}
with $P_J(x)$ the $J$th order of Legendre polynomial. Being an elastic
equal-mass scattering process, the analytical expressions of the
partial-wave $\pi\pi$ scattering amplitudes can be easily evaluated
and they take the neat forms
\begin{align} {T_0^0}_{tree}^{(2)}=&\frac{2s-m_{\pi}^2}{2F_{\pi}^2},\\
{T_0^0}_{tree}^{(4)}=&\frac{5l_3^r m_{\pi}^4}{2F_{\pi}^2}-\frac{l_4^r(m_{\pi}^4-2m_{\pi}^2s)}{F_{\pi}^4}
+\frac{l_2^r(28m_{\pi}^4-20m_{\pi}^2+7s^2)}{3F_{\pi}^4}+\frac{l_1^r(44m_{\pi}^4-40m_{\pi}^2+11s^2)}{3F_{\pi}^4},\\
{T_0^0}_{loop}^{(4)}=&\frac{1}{1728F_{\pi}^4{\pi}}(613m_{\pi}^4-440m_{\pi}^2s+142s^2-3(119m_{\pi}^4-88m_{\pi}^2s+50s^2)\ln\left[\frac{m_{\pi}^2}{\mu^2}\right]\notag\\
&+27\sqrt{1-\frac{4m_{\pi}^2}{s}}(m_{\pi}^2-2s)^2\ln\left[-\frac{1}{2m_{\pi}^2}\left(-2m_{\pi}^2+s-\sqrt{s(-4m_{\pi}^2+s)}\right)\right])\notag\\
&+\frac{1}{1728F_{\pi}^4{\pi}(4m_{\pi}^2-s)}((4m_{\pi}^2-s)(506m_{\pi}^4-130m_{\pi}^2s+11s^2)\notag\\
&-6\sqrt{s(-4m_{\pi}^2+s)}(75m_{\pi}^4-40m_{\pi}^2s+7s^2)\ln\left[\frac{1}{2m_{\pi}^2}\left(-2m_{\pi}^2+s-\sqrt{s(-4m_{\pi}^2+s)}\right)\right]\notag\\
&+18(25m_{\pi}^6-6m_{\pi}^4s)\ln[\frac{1}{2m_{\pi}^2}\left(-2m_{\pi}^2+s-\sqrt{s(-4m_{\pi}^2+s)}\right)]^2),
\end{align}
for the $(I,J)=(0,0)$ case. The analytical results for the $(I,J)=(1,1)$ are found to be
\begin{align}
{T_1^1}_{tree}^{(2)}=&\frac{s-4m_{\pi}^2}{6F_{\pi}^2},\\
{T_1^1}_{tree}^{(4)}=&\frac{l_1^r(4m_{\pi}^2-s)s}{3F_{\pi}^4}+\frac{l_2^r(-4m_{\pi}^2+s)s}{6F_{\pi}^4}
+\frac{l_4^r(-4m_{\pi}^2+m_{\pi}^2s)}{3F_{\pi}^4},\\
{T_1^1}_{loop}^{(2)}=&\frac{(-4m_{\pi}^2+s)(s-12m_{\pi}^2\ln\left[\frac{m_{\pi}^2}{\mu^2}\right])}{576F_{\pi}^4\pi}+\frac{\sqrt{1-\frac{4m_{\pi}^2}{s}}(-4m_{\pi}^2+s)^2\ln\left[-\frac{-2m_{\pi}^2+s-\sqrt{s(-4m_{\pi}^2+s)}}{2m_{\pi}^2}\right]}{576F_{\pi}^4\pi}\notag\\
&+(-\frac{1}{1728F_{\pi}^4\pi(-4m_{\pi}^2+s)^2}(-480m_{\pi}^8+716m_{\pi}^6s-297m_{\pi}^4s^2+33m_{\pi}^2s^3+s^4+3\sqrt{s(-4m_{\pi}^2+s)}\notag\\
&(-36m_{\pi}^6+72m_{\pi}^4s-16m_{\pi}^2s^2+s^3)\ln\left[\frac{-2m_{\pi}^2+s-\sqrt{s(-4m_{\pi}^2+s)}}{2m_{\pi}^2}\right]\notag\\
&+18(6m_{\pi}^8+13m_{\pi}^6s-3m_{\pi}^4s^2)\ln\left[\frac{-2m_{\pi}^2+s-\sqrt{s(-4m_{\pi}^2+s)}}{2m_{\pi}^2}\right]^2)).
\end{align}
The corresponding expressions in the $(I,J)=(2,0)$ case are given by
\begin{align}
{T_2^0}_{tree}^{(2)}=&\frac{2m_{\pi}^2-s}{2F_{\pi}^2},\\
{T_2^0}_{tree}^{(4)}=&\frac{2l_3^r m_{\pi}^4}{F_{\pi}^4}-\frac{l_4^r(2m_{\pi}^4-m_{\pi}^2s)}{F_{\pi}^4}
+\frac{l_1^r(4m_{\pi}^4-2m_{\pi}^2+s^2)}{3F_{\pi}^4}+\frac{l_2^r(8m_{\pi}^4-7m_{\pi}^2+2s^2)}{3F_{\pi}^4},\\
{T_2^0}_{loop}^{(2)}=&\frac{1}{1728F_{\pi}^4{\pi}}(274m_{\pi}^4-212m_{\pi}^2s+64s^2-6(67m_{\pi}^4-50m_{\pi}^2s+10s^2)\ln\left[\frac{m_{\pi}^2}{\mu^2}\right]\notag\\
&+27\sqrt{1-\frac{4m_{\pi}^2}{s}}(-2m_{\pi}^2+s)^2\ln\left[-\frac{-2m_{\pi}^2+s-\sqrt{s(-4m_{\pi}^2+s)}}{2m_{\pi}^2}\right])+(\frac{1}{3456F_{\pi}^4\pi(4m_{\pi}^2-s)}\notag\\
&(-1232m_{\pi}^6+540m_{\pi}^4s+42m_{\pi}^2s^2-25s^3-6\sqrt{s(-4m_{\pi}^2+s)}(6m_{\pi}^4-32m_{\pi}^2s+11s^2)\notag\\
&\ln\left[\frac{-2m_{\pi}^2+s-\sqrt{s(-4m_{\pi}^2+s)}}{2m_{\pi}^2}\right]+36(m_{\pi}^6+3m_{\pi}^4s)\ln\left[\frac{-2m_{\pi}^2+s-\sqrt{s(-4m_{\pi}^2+s)}}{2m_{\pi}^2}\right]^2))\,.
\end{align}.

We will take
\begin{align}
T_{{\chi PT},IJ}=&\frac{1}{16\pi}\left({T_J^I}_{tree}^{(2)}+{T_J^I}_{tree}^{(4)}+{T_J^I}_{loop}^{(2)}\right),
\end{align}
in our analyses.

The values of the $SU(2)$ LECs are taken from the updated studies in Ref.~\cite{Bijnens:2014lea}. The renormalized $l_i^r$ take the form
\begin{align}
    l_i^r=\frac{\gamma_i}{32\pi^2}\left(\bar{l_i}+ln\left[\frac{m_{\pi}^2}{\mu^2}\right]\right),
    (\gamma_1=\frac{1}{3},\gamma_2=\frac{2}{3},\gamma_3=-\frac{1}{2},\gamma_4=2 )
\end{align}
with
\begin{align}
\bar{l_1}=-0.4,~\bar{l_2}=4.3,~\bar{l_3}=2.9,~\bar{l_4}=4.4,~F=85.8\mathrm{MeV},~\mu=770\mathrm{MeV}.
\end{align}
The renormalized pion decay constant $F_\pi$ is given by
\begin{align}
F_{\pi}=F\left(1+l_4^r\frac{m_{\pi}^2}{F^2}-\frac{1}{16\pi^2}\ln\left[\frac{m_{\pi}^2}{\mu^2}\right]\frac{m_{\pi}^2}{F^2}\right)\,.
\end{align}

Thus, the discontinuity of the $l.h.c.$ integral in eq.~(\ref{eq.lhcf}) can be expressed by the $\chi$PT amplitude as
\begin{align}
&\text{Im}_{L}f(s)=-\frac{1}{2\rho(s)}\ln|S_{\chi PT}(s)|,\\
&S_{\chi PT}(s)=1+2i\rho(s)T_{\chi PT}(s),
\end{align}
with the indices of the isospin and the angular momentum omitted. To suppress the improper behavior of the perturbative $\chi$PT amplitudes in high energy region, a cutoff parameter $\Lambda_{L}$ is introduced in the numerical analysis. The $l.h.c.$ integral is explicitly written as
\begin{align}\label{eq.lhcf2}
f(s)=-\frac{s}{\pi}\int_{-\Lambda_{L}}^{0}\frac{\ln|1+2i\rho(s')T_{\chi PT}(s')|}{2\rho(s')s'(s'-s)}ds'\,.
\end{align}

In the elastic scattering region, the partial wave $S$-matrix is
parameterized as $S(s)=e^{2i\delta(s)}$ where the indices of isospin
and the angular momentum are suppressed for simplicity. Thus, the
contributions of all poles and cuts to the phase shifts are additive, $i.e.$,
\begin{align}
\delta(s)=\sum_{i}\delta_{i}^{pole}(s)+\delta_{BG},
\end{align}
where $\delta^{pole}$ represents the contributions from the various poles, and the background contribution from the $l.h.c.$ takes the form
\begin{align}
\delta_{BG}(s)=\rho(s)f(s)\,.
\end{align}

\subsection{crossing symmetry of $\pi\pi$ scattering and BNR relations}\label{bnr}

Crossing symmetry demands that the $\pi\pi$ scattering amplitudes of
$s$, $t$, and $u$ channels could be expressed by the same analytic
functions in different physical regions, which can not be
automatically fulfilled by the PKU representation. To remedy this
deficiency, we use the BNR relations~\cite{Balachandran:1968zza,Roskies:1969pe}, which are the relations about different partial-wave amplitudes of $\pi\pi$ scattering, as additional constraints in the fits. In this work, since there are only
lattice data for $IJ=00, 11, 20$ channels, we only need to study
the following five BNR relations concerning $s$ and $p$-waves, which are given by
\begin{align}
\text{\uppercase\expandafter{\romannumeral1}}&:\qquad\int_{0}^{4m^2_\pi}(s-4m^2_\pi)(3s-4m^2_\pi)[t_{0}^{0}(s)+2t_{0}^{2}(s)]ds=0,\notag\\
\text{\uppercase\expandafter{\romannumeral2}}&:\qquad\int_{0}^{4m^2_\pi}(s-4m^2_\pi)R_{0}^{0}[2t_{0}^{0}(s)-5t_{0}^{2}(s)]ds=0,\notag\\
\text{\uppercase\expandafter{\romannumeral3}}&:\qquad\int_{0}^{4m^2_\pi}(s-4m^2_\pi)R_{1}^{0}[2t_{0}^{0}(s)-5t_{0}^{2}(s)]ds-9\int_{0}^{4m^2_\pi}(s-4m^2_\pi)^{2}R_{0}^{1}t_{1}^{1}ds=0,\notag\\
\text{\uppercase\expandafter{\romannumeral4}}&:\qquad\int_{0}^{4m^2_\pi}(s-4m^2_\pi)R_{2}^{0}[2t_{0}^{0}(s)-5t_{0}^{2}(s)]ds+6\int_{0}^{4m^2_\pi}(s-4m^2_\pi)^{2}R_{1}^{1}t_{1}^{1}ds=0,\notag\\
\text{\uppercase\expandafter{\romannumeral5}}&:\qquad\int_{0}^{4m^2_\pi}(s-4m^2_\pi)R_{3}^{0}[2t_{0}^{0}(s)-5t_{0}^{2}(s)]ds-15\int_{0}^{4m^2_\pi}(s-4m^2_\pi)^{2}R_{2}^{1}t_{1}^{1}ds=0,
\end{align}
where the functions $R_i^j$s are respectively defined as
\begin{align}
&R_0^0=1~,~R_0^1=1~,\notag\\
&R_1^0=3s-4m^2_\pi~,~R_1^1=5s-4m^2_\pi~,\notag\\
&R_2^0=10s^2-32s m^2_\pi +16m^4_\pi~,~R_2^1=21s^2-48s m^2_\pi+16m^4_\pi~,\notag\\
&R_3^0=35s^3-180s^2m^2_\pi+240sm^4_\pi-64m^6_\pi~.
\end{align}
These BNR relations can be rewritten in a concise form
\bqa
\int_0^{4m^2_\pi}ds[P_{00}^i(s)T_0^0(s)+P_{11}^i(s)T_1^1(s)+P_{20}^i(s)T_0^2(s)]=0,
\eqa
where $i$ ranges from I to V and $P_{IJ}^i$ are polynomials for
different channels. These relations should be
respected precisely if the partial wave amplitudes exactly obey the crossing
symmetries. It should be noted that not only the explicit $m_\pi$ factors but also the scattering amplitudes in the BNR relations vary according to the pion masses for different sets of lattice QCD data. It is interesting to investigate how these BNR relations are satisfied at different values of pion masses. This may provide us a different insight about the crossing symmetry as a function of $m_\pi$.

\section{Numerical analyses and discussions}\label{numerical}

Our analysis is based on the two sets of lattice simulation data from
Hadron Spectrum Collaboration with $m_\pi=$391 MeV~(referred to as
Data391 here) and $m_\pi=$236 MeV~(referred to as Data236 here).
Data391 includes the $IJ=00$ $\pi\pi$ scattering phase shift in
ref.~\cite{Briceno:2016mjc}, the one for $IJ=11$ in
ref.~\cite{dudek:PhysRevD.87.034505,dudek:PhysRevD.90.099902}, and the
one for the
$IJ=20$ in ref.~\cite{Dudek:2012gj}. Data236 includes the $IJ=00$
$\pi\pi$ scattering phase shift in ref.~\cite{Briceno:2016mjc} and
the one for  the
$IJ=11$ in ref.~\cite{Wilson:2015dqa}. The
$IJ=20$ phase of $m_\pi=$391 MeV exhibits a similar behavior to the
experimental ones, therefore one would expect rather mild deviations of the phase shifts with $m_\pi=236$~MeV in this channel, comparing with the physical values.

A few remarks about the numerical procedure are needed.  In the
numerical analysis, we first perform separate fits to each set of data
of different quantum numbers using the PKU representation.
Then, a combined fit to all the three channels with $IJ=00, 20, 11$, together
with the BNR relations introduced as the penalty functions to meet the
requirements of crossing symmetry, is performed.

It is noted that we will simultaneously take into account both the uncertainties of the phase shifts and also the error bars of scattering three-momentum squared provided by the lattice calculations. To be more specific, we will take the average for each phase shift at a given energy point within its uncertainty range when performing the fits of the lattice data.

\subsection{Individual fit of the $IJ=20$ channel}

\begin{figure}[hbtp]
\centering
\includegraphics[width=5.5cm]{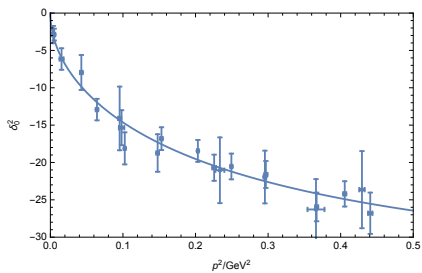}
\caption{The fitted $\pi\pi$ phase shift of $m_\pi=391\ \mathrm{MeV}$ in $IJ=20$ compared with the lattice data in ref. \cite{Dudek:2012gj}.    }\label{fig.IJ20only.391}
\end{figure}

According to the experimental data~\cite{Hoogland:1974cv}, the $\pi\pi$ scattering in the isotensor channel has no resonant structure in the low-energy region, so the phase shifts are negative and decrease smoothly from the threshold. The lattice simulation with
$m_\pi=391\ \mathrm{MeV}$ shows a similar behavior below the $4\pi$
threshold and the data could be well described by the background
contribution in our study, which is determined from the $l.h.c.$ integral as shown
in Fig.~\ref{fig.IJ20only.391}. However, it is worth mentioning that
in later discussion of  the combined fit the $l.h.c$ effect is found
to be inefficient to describe the the $IJ=20$ lattice data. It turns
out that a virtual state pole is needed in this channel,  and is located at about
$s_0=0.049m^2_\pi$ on the real axis of the second Riemann sheet of the
complex $s$-plane~\cite{Ang:2001bd}. A virtual state pole contributes
a mildly rising positive phase shift above the threshold, which slightly
compensates the contribution from the background. Using
only the $IJ=20$ data could not separate the the virtual state
contribution from the background so that the virtual pole position and
the cutoff parameter could not be determined. The combined fit with
constraints {of BNR relation} from other channels turns out to
be crucial in determining those parameters.

\subsection{Individual fit of the $IJ=11$ channel}

The $IJ=11$ phase shifts of lattice simulation both at $m_\pi=391$ MeV and $m_\pi=236$ MeV present typical resonance structures. With a simple $K$-matrix parameterization fit of Data391, the mass and width of the $\rho$ are $854.1\pm 1.1$ MeV and $12.4\pm 0.6$ MeV, respectively, for $m_\pi=391\ \mathrm{MeV}$ in Ref.~\cite{dudek:PhysRevD.87.034505,dudek:PhysRevD.90.099902}, while those of $\rho$ from Data236 are $783\pm 2$ MeV and $90\pm 8$ MeV, respectively, for $m_\pi=236\ \mathrm{MeV}$ in Ref.~\cite{Wilson:2015dqa}.

\begin{figure}[htbp]
\centering
\includegraphics[width=5.5cm]{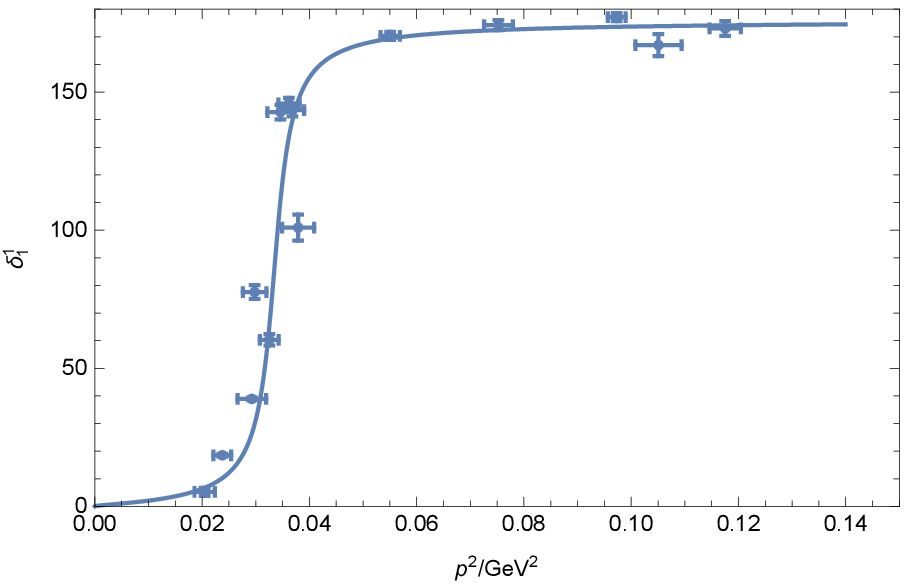}
\includegraphics[width=5.5cm]{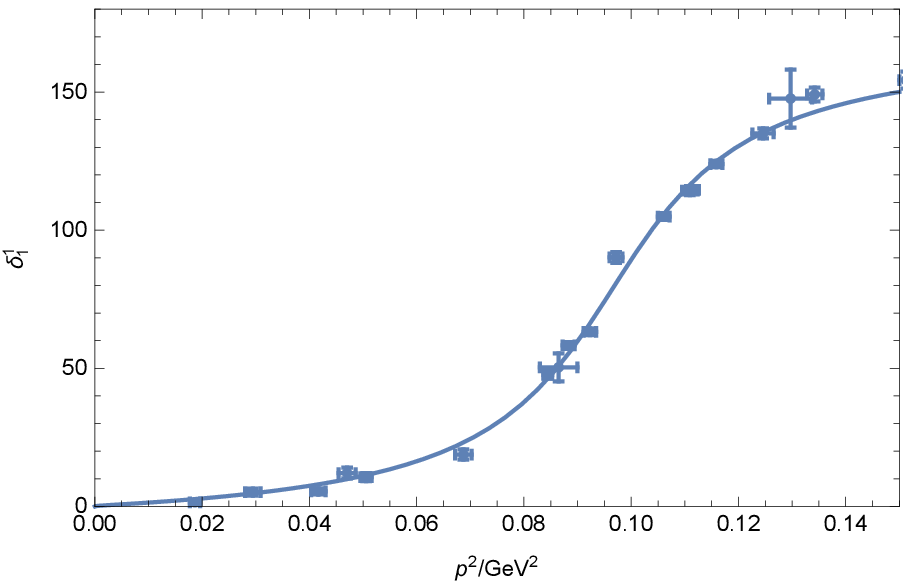}
\caption{The $IJ=11$ $\pi\pi$ phase shifts of $m_\pi=391\ \mathrm{MeV}$~(left) and $m_\pi=236\ \mathrm{MeV}$~(right), respectively, in $IJ=11$ compared with the lattice data in refs. \cite{dudek:PhysRevD.87.034505,dudek:PhysRevD.90.099902}.}\label{fig.IJ11only}
\end{figure}

The fit quality of the lattice phase shifts is quite good in our study, as shown in Fig.~\ref{fig.IJ11only}. The resulting pole positions of the $\rho$ are determined to be $\sqrt{s_\rho}=862.9-\frac{10.1}{2}i$ MeV for $m_\pi=391$ MeV and at $\sqrt{s_\rho}=782.2-\frac{96.3}{2}i$ MeV for $m_\pi=236$ MeV, which agree well with the results from Refs.~\cite{dudek:PhysRevD.87.034505,dudek:PhysRevD.90.099902,Wilson:2015dqa}.

\subsection{Individual fit of the $IJ=00$ channel}

\begin{figure}[h]
\centering
\includegraphics[width=5.5cm]{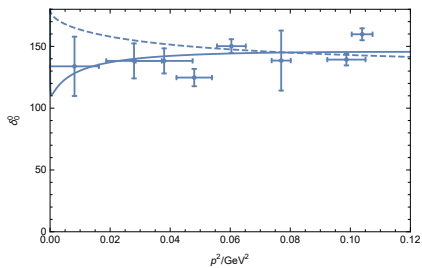}
\caption{The fitted $\pi\pi$ phase shift of $IJ=00$ with a bound-state pole~(dashed) or a bound-state pole and a virtual-state pole~(solid), compared with data in ref.~\cite{Briceno:2016mjc}, when $m_\pi =391$ MeV.}\label{fitIJ00only}
\end{figure}

Model-independent analyses demonstrate that the $\sigma$ resonance,
which corresponds to a pair of complex conjugate poles  deep in the complex $s$-plane on the second Riemann sheet,  must exist
to reproduce the observed experimental phase
shifts~\cite{Xiao:2000kx,Zhou:2004ms,Caprini:2005zr,Garcia-Martin:2011nna,Pelaez:2015qba}.
The lattice phase shifts with $m_\pi=236$~MeV also feature
the contribution of a broad resonance, which resembles the behavior
observed in   the experiments, even though the pole positions
determined in different parameterization forms may vary. In contrast,
the lattice phase shifts with $m_\pi=391\ \mathrm{MeV}$
decrease rapidly from $180^\circ$ at the threshold to about $130^\circ$ and
then approach almost to a constant value within large energy ranges \footnote{The
lattice data in ref.~\cite{Briceno:2016mjc} define the phase at
$\pi\pi$ threshold to be $180^\circ$.}, typically indicating the
existence of a bound state. We first make a simple
fit to the lattice data by assuming that it is contributed by a bound-state
pole and the left-hand cut.
The quality of such a tentative fit is quite poor with $\chi^2=32.2$ as shown by the dashed line in Fig.~\ref{fitIJ00only}, even though there are only 8 data points in this channel. The bound-state pole is located at $\sqrt{s_b}=590$ MeV and the cutoff parameter $\Lambda_L$ turns out to be vanishing. We verify that the contribution of the $l.h.c.$ integral drives further away of the phase shifts from the lattice results. This clearly indicates the deficiency by only including the contribution of a bound-state pole. It is noticed that the difference between such a fit result and the lattice data could be reduced by a gradually increasing phase
shift from the threshold, and this is exactly the effect from a virtual-state pole. Therefore, an efficient way to obtain a reasonable fit is to introduce a virtual-state pole into the formalism of Eq.~\eqref{eq.smat}.

The fit by including a bound-state pole as well as a virtual-state pole significantly
improves the fit quality with $\chi^2=17.0$. The bound state pole is determined to be $\sqrt{s_b}=781$ MeV and the virtual-state pole is $\sqrt{s_v}=709$ MeV. The reproduction of the lattice phase shifts with both bound- and virtual-state poles is given by the solid line in Fig.~\ref{fitIJ00only}.
It is the cancellation between the contributions from the bound-state pole
and the virtual-state pole that causes a sharp decrease of phase shifts near
the $\pi\pi$ threshold. This improvement suggests that the existence
of the nearby virtual-state pole in the $IJ=00$ channel is greatly helpful in
describing the lattice data. Such an improvement could be more obvious if we modify the Data391 by reducing uncertainties of the first three data points in $IJ=00$ to one-fourth of their own  values respectively~(referred to modified Data391). The $\chi^2$ of fit of modified Data391 without a virtual state pole is 70.0, while that with a virtual-state pole is 17.9.

The description of the $\sigma$ at $m_\pi=391$~MeV as a pair of bound and virtual poles is a novel finding in our study, and to our knowledge it has not been reported in the literature or by other dispersive technique as in
Ref.~\cite{Danilkin:2020pak}.

The situation for the $\sigma$ at the physical pion mass seems a bit different, since
it is found that the broad $\sigma$ resonance could be easily generated by only including
the nonperturbative two-pion interactions.
Recall that in the Friedrichs-Lee model, which devotes to the explanation of all the
low-lying scalar nonet below 1.0 GeV and the scalar nonet higher than
1.0 GeV, the dynamically generated $\sigma$ poles on the complex
$s$-plane will move towards the real axis and become two different
virtual state poles, and then become a bound state pole and a virtual state
pole as the coupling strength
increases~\cite{Zhou:2020vnz,Zhou:2020moj}. The bound-virtual states
scenario also happens when the quark mass become large.
Such a pole behavior of the $\sigma$ resonance has also been noticed in
the calculations by unitarizing $\chi$PT amplitude with a varying quark
mass $m_q$~\cite{Pelaez:2015qba}. A resonance is described by a
pair of conjugated pole on the unphysical complex $s$-plane as
required by the Schwartz reflection rule. If the two poles meet each
other on the real axis when the coupling strength or other parameter
changes, they could not disappear but become two virtual poles or a
pole pair of a bound state and a virtual
state~\cite{Hyodo:2014bda,Hanhart:2014ssa}.
For an elementary bound state, the virtual-state pole is typically
close to the bound-state pole when the coupling is weak.
While for a molecular-type bound state,  which is an essentially
nonperturbative phenomenon, there is just one bound-state pole around the interested energy region and the virtual-state pole usually lies far away from the bound-state pole.

\subsection{The combined fit with constraints from crossing symmetry}

We have performed fits for the $IJ=20$, $11$, and $00$ channels separately and the outcome of each channel has been discussed in the previous sections. The
PKU representation respects the unitarity and analyticity of partial
wave amplitudes by construction, but the crossing symmetry of $\pi\pi$ scattering needs
to be restored by the BNR relations. To impose the crossing
symmetry on the partial wave amplitude in PKU representation, we define the penalty functions of
BNR relations as follows:
\bqa\label{penalty}
\chi^2_{BNR}=\frac{1}{\epsilon^2}\sum_{i=I}^{V}\frac{|\int_0^{4m^2_\pi}ds[P_{00}^i(s)T_0^0(s)+P_{11}^i(s)T_1^1(s)+P_{20}^i(s)T_0^2(s)]|^2}{\{\int_0^{4m^2_\pi}ds[|P_{00}^i(s)T_0^0(s)|+|P_{11}^i(s)T_1^1(s)|+|P_{20}^i(s)T_0^2(s)|]\}^2}
\eqa
and put it into the total $\chi^2$ as
\bqa
\chi^2_{tot}=\chi^2_{00}+\chi^2_{11}+\chi^2_{20}+\chi^2_{BNR}.
\eqa
The penalty factor $\epsilon^{-2}$ is chosen at $10^{-4}$ to
ensure that the violation of the BNR relations is at the order of
about several percent.

\begin{figure}[th]
\centering
\includegraphics[width=5.5cm]{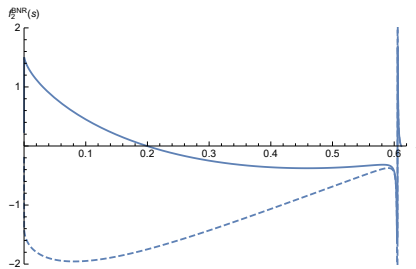}
\caption{The integrand function of the second BNR relation, $f_2^{BNR}(s)$, within the region of $0<s<4m_\pi^2$~($m_{\pi}=391$ MeV) when the virtual state pole of $IJ=20$ channel exists~(solid) or does not exist~(dashed).}\label{figbnr2}
\end{figure}

Nevertheless, such a combined fit might suffer  from the fact
that the uncertainties of the lattice data in different channels are
rather different. In particular, the error bars of the data in the
$IJ=11$ channel are around one order of magnitude smaller than those in the other channels.
To enforce the importance of $IJ=00$ channel and to emphasize the
effect of the data near threshold, the modified Data391 is used in the
fit. In the fit it is found that the cutoff parameters of three channels can not be constrained at the same value, and the best fit of $\Lambda^2_{IJ=20}$ is larger than $\Lambda^2_{IJ=00}$ and $\Lambda^2_{IJ=11}$. One should remember that there exist higher resonances and right hand cuts in both $IJ=00$ and $11$ channels which are unable to be counted in this analysis. Their contributions are all positive and might be absorbed in the numerical fit of the left-hand cuts. The final fit results are in the follows:
\begin{align}
&\chi_{tot}^2=\chi_{00}^2+\chi_{11}^2+\chi_{20}^2+\chi_{BNR}^2=30.0+1154.3+20.4+282.1~;\notag\\
&\sqrt{s_{b,IJ=00}}=774\pm 6\mathrm{MeV}~,~\sqrt{s_{v,IJ=00}}=716\pm 28\mathrm{MeV}~~,\notag\\
&M_{\rho}=863.2\pm 0.6\mathrm{MeV}~,~\Gamma_{\rho}=11.5\pm 1.2\mathrm{MeV}~,\notag\\
&\sqrt{s_{v,IJ=20}}=247\pm 99\mathrm{MeV}~.
\end{align}
We have also tried the combined fit without including the virtual state pole in $IJ=20$ for comparison. It turns out that, in such a
scenario, the second BNR relation will be severely violated because
the  integrand,
$f^{BNR}_2(s)=(s-4m^2_\pi)R_{0}^{0}[2t_{0}^{0}(s)-5t_{0}^{2}(s)]$, is
almost always negative between 0 and $4m^2_\pi$, as shown in fig.\ref{figbnr2}, such that the integration can
not be cancelled to fulfill the BNR constraint. This indicates that the crossing symmetry with unphysically large pion masses dictates the existence of the virtual-state pole in the $IJ=20$ channel. It is noted that in the physical case the existence of such a virtual-state pole is also advocated in Ref.~\cite{Ang:2001bd}.

\begin{figure}[H]
\centering
\includegraphics[width=5.5cm]{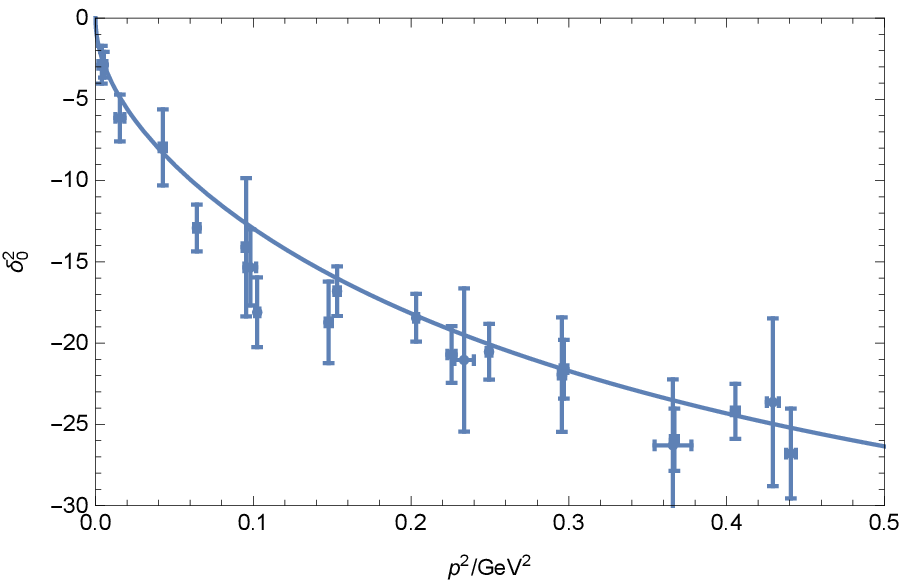}
\includegraphics[width=5.5cm]{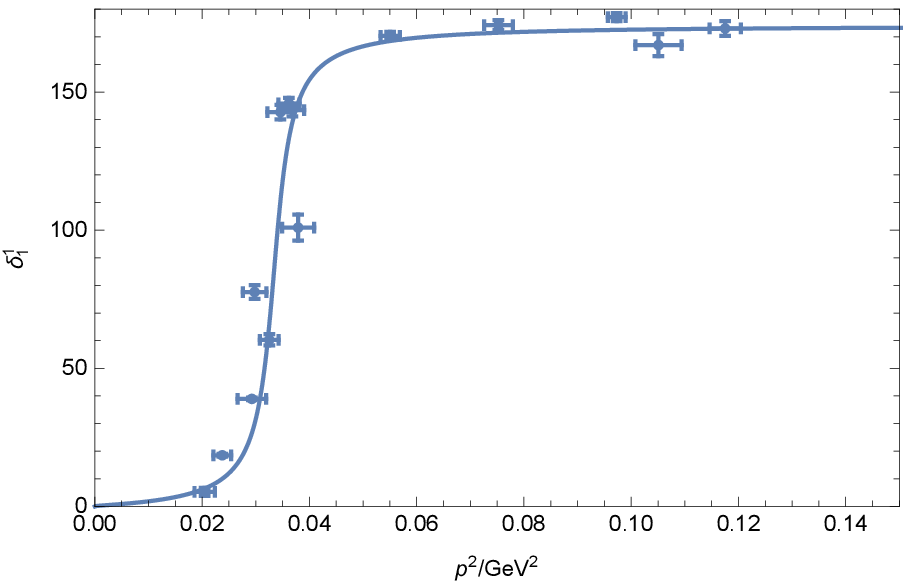}
\includegraphics[width=5.5cm]{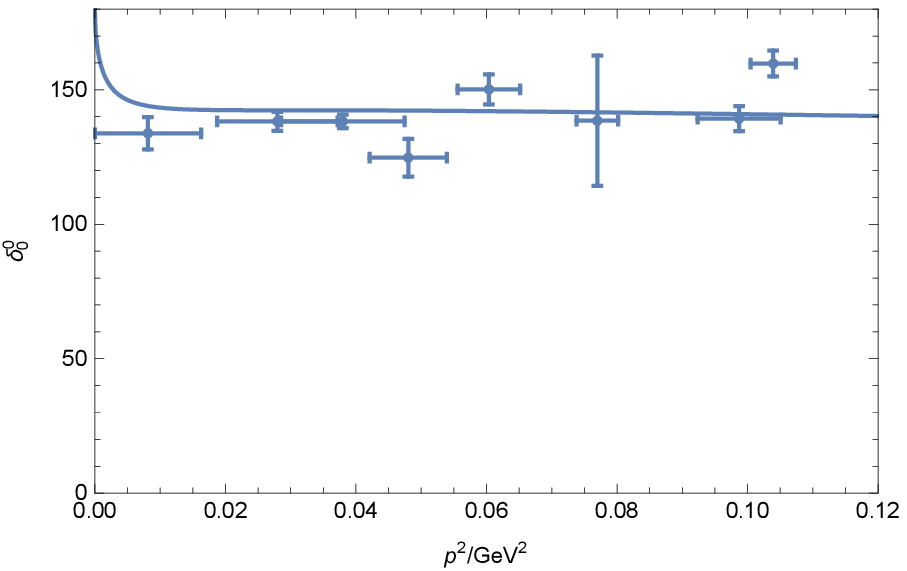}
\caption{Fit results of modified lattice data of $m_\pi=391\mathrm{MeV}$.}
\end{figure}

\begin{figure}[H]
\centering
\includegraphics[width=5.5cm]{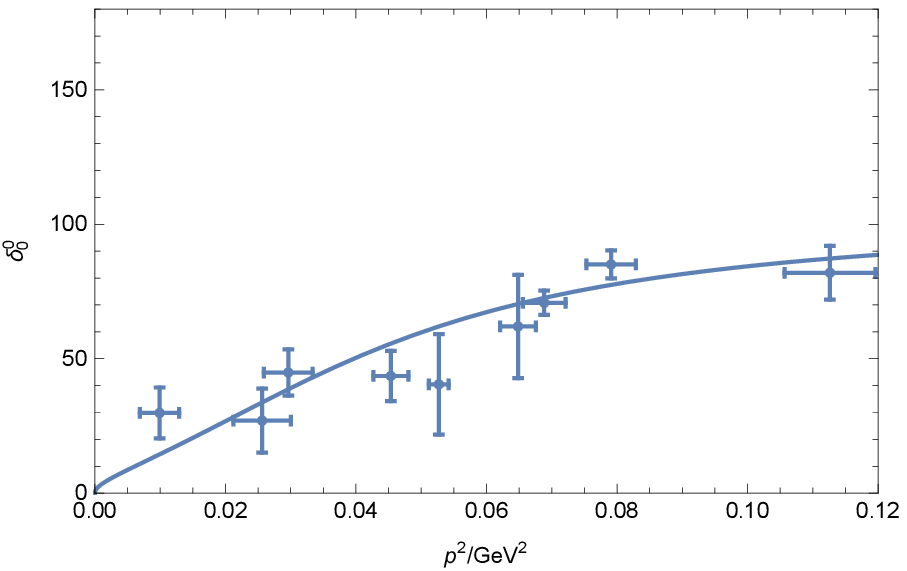}
\includegraphics[width=5.5cm]{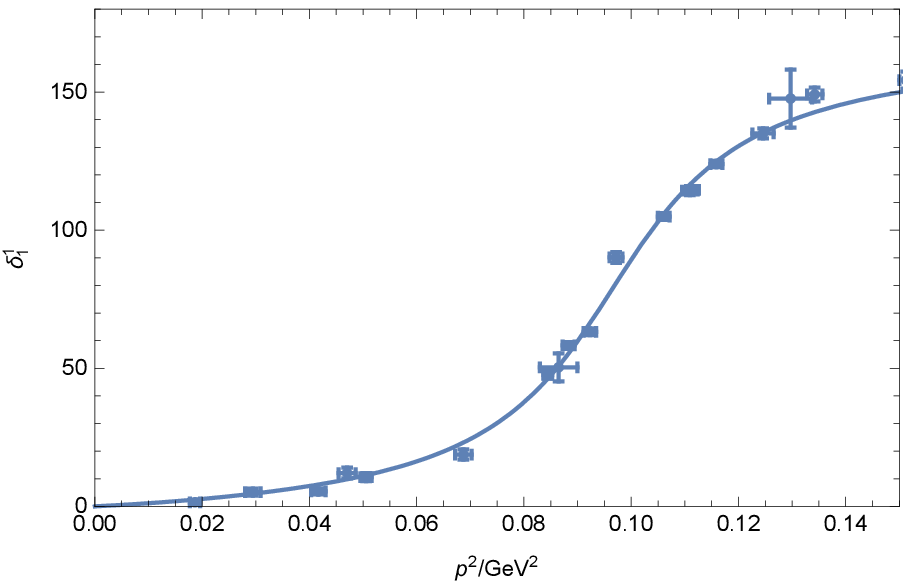}
\caption{Fit results of lattice data of $m_\pi=\mathrm{236MeV}$.}\label{fig.236}
\end{figure}

For the case of $m_{\pi}$=236MeV, we take a simultaneous fit of the
lattice blue data in the $IJ=00$ and $IJ=11$ channels, due to the absence of the data sets in the $IJ=20$ channel, as shown in Fig.~\ref{fig.236}, without implementing the constraints of the BNR relations. The fit parameters are shown in the following:
\begin{align}
&\chi_{tot}^2=\chi_{00}^2+\chi_{11}^2=8.8+82.0~,\notag\\
&M_{\sigma}=610\pm 11\mathrm{MeV}~,~\Gamma_{\sigma}=327\pm 8\mathrm{MeV}~,\notag\\
&M_{\rho}=782\pm 2\mathrm{MeV}~,~\Gamma_{\rho}=96\pm 4\mathrm{MeV}.
\end{align}

This result shows that when $m_\pi=236$ MeV the $\sigma$ pole corresponds a broad resonance located on the second sheet of the complex $s$-plane but closer to the real axis compared with the physical one, which is at about $\sqrt{s_{\sigma}}=470 - i\frac{570}{2}$ MeV
in ref.~\cite{Zhou:2004ms}. The $\rho$ resonance also becomes narrower
than the physical $\rho$ pole at $\sqrt{s_{\rho}}=757 - i\frac{152}{2}$ MeV.

\begin{figure}[H]
\centering
\includegraphics[width=5.5cm]{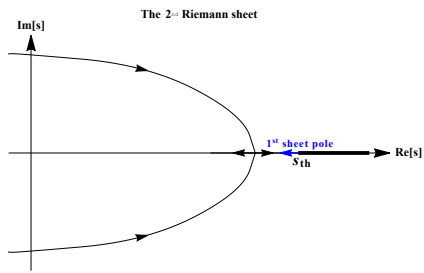}
\caption{{A rough qualitative picture of} the $\sigma$ pole trajectory  as the pion mass increases. The thick line starting from $s_{th}$ represents the unitarity cut. The blue arrow and text represents the trajectory of the bound state pole which comes from one of the virtual state pole across the threshold.}\label{poletrajectory}
\end{figure}

Considering  the fit results of lattice data with $m_\pi=391\ \mathrm{MeV}$
and $m_\pi=236\ \mathrm{MeV}$ and the fit result of the physical
scattering phase shifts together, a picture of $\sigma$ pole trajectory
under the change of pion mass could be imagined. The $\sigma$
resonance corresponds to a pair of conjugated poles in the deep complex
$s$-plane on the second Riemann sheet at the physical pion mass.
As the pion mass increases, the conjugated poles will move
towards the real $s$-axis with their imaginary parts becoming smaller and
smaller. When
the pion mass becomes large enough, the pair of poles will collide with
each other on the real axis and become two virtual-state poles. As the
pion mass keeps increasing, one of the virtual state pole will move
down along the real axis and the other one moves up across the
threshold to the first Riemann sheet and becomes a bound state, {similar to the results in} the
Relativistic Friedrichs-Lee-QPC scheme or unitarized
$\chi$PT~\cite{Zhou:2020vnz,Pelaez:2015qba,Hanhart:2014ssa} (see Fig.\ref{poletrajectory}).

\section{summary}

In this paper, we analyze the lattice QCD data of $\pi\pi$ scattering
using a novel dispersive solution of the S-matrix, i.e., the PKU representation, with the constraints of crossing symmetry
implemented by the BNR relations. This scheme has the  merits of
respecting the unitarity, analyticity, and crossing symmetry of
scattering amplitudes, and the contribution from the left-hand cut is
estimated using the $SU(2)$ $\chi$PT calculation.

It is evident in our analysis that the lattice data at $m_\pi$=391 MeV in the $IJ=00$ channel
obviously prefer the scenario of a bound state pole and a virtual
state pole for the $\sigma$ rather than the
one of just one bound state pole.
The bound state pole is located close to
the $\pi\pi$ threshold while the virtual state pole is  a bit farther away. However,
the virtual pole can not be determined precisely due to the large
uncertainty of the phase shift data just above the threshold. This
bound-virtual state pair phenomenon could also be
  seen in the Friedrichs-Lee-QPC scheme or
unitarizing the $\chi$PT  where the pair is dynamically generated. The existence of the virtual state pole may
also provide more insight into our understanding of the enigmatic
$\sigma$ resonance. A general picture of the pole trajectory
with varying pion mass can now be figured out. Furthermore, the combined fit with the constraints
of crossing symmetry also provide further support of the existence of
another virtual state pole in the $IJ=20$ $\pi\pi$ channel.

Our study shows that the lattice QCD simulation is able to supply more information about the nature of the light scalar resonances even though the pion mass is unphysically large. It provides another variation of the parameters which could not be tuned in the physical world.  More precise data near the threshold in the lattice simulation are suggested for extracting the precise pole position of $\sigma$ and other resonances. With such precise lattice data, it is also fascinating to further investigate the crossing symmetry in two-meson scattering at different quark masses.

\begin{acknowledgments}
This work is supported by National Science Foundation of China under contract Nos. 11975075, 11575177, 11975090 and 12150013. This work is also supported by “the Fundamental Research Funds for the Central Universities”.
\end{acknowledgments}


\bibliography{Ref}

\end{document}